\documentclass[12pt]{article}
\usepackage{cite}
\usepackage{amsmath}
\usepackage{graphicx}
\title{\bf  Analytical  models of finite thin disks  in a  magnetic
field  }

\author{E Cardona-Rueda and G Garc\'{\i}a-Reyes\thanks{
Corresponding Author, E-mail: ggarcia@utp.edu.co}  \\
Universidad  Tecnol\'ogica de Pereira, Departamento de
F\'{\i}sica,  Pereira, Colombia }

\date{ }

\begin{document}

\maketitle

\begin{abstract}
Analytical  models of  axially symmetric    thin disks of finite extension in presence of
magnetic field are presented   based on the well-known  Morgan-Morgan solutions.
The source of the magnetic field is cons\-tructed
separating the equation corresponding to the Ampere's law of electrodynamics
in spheroidal  oblate coordinates.
This produces  two asso\-ciated Legendre equations of first  order for the
magnetic potential  and hence   that  can be expressed as a series of
associated  Legendre functions of the same order.
The discontinuity of its normal derivate   across the disk
allows us  to interpret   the source of the magnetic field as a ringlike current
distribution extended  on all the plane of the disk.  We also study  the
circular speed curves  or rotation curve for
equatorial circular orbits of charged test particles both inside and outside the disk.
The stability of the  orbits is analyzed for radial perturbation
using a extension of the Rayleigh criterion.

 \end{abstract}

\maketitle

\newpage

\section*{1. Introduction}
Axially symmetric   thin disks  are
important in astrophysics as models   of  galaxies,  accretion disks
and certain stars.  In the case of galaxies it is motivated by the fact that the
main part of the mass  of the galaxies is concentrated in
the galactic disk  \cite{Binney}.  Even though a realistic disk  has thickness,
in first approximation these astrophysical objects can
be considered to be very thin, e.g., in our Galaxy the radius of the disk is
10 kpc and its thickness is 1 kpc.
Thin disks in presence of magnetic field are also of astrophysical importance.
In fact, there is strong observational evidence that magnetic fields
are present in
all galaxies and galaxy clusters \cite{ref1,ref2, Klein}. 
These fields are characterized by a  strength of the order of  $\mu$ G.
For example, the strength of the total magnetic field in our
Milky Way from radio synchrotron measurements is about $6 \mu $ G  \cite{Beck}.
Observational data  also provide   a strong
support of the existence of extragalactic magnetic fields of
at least $B \simeq 10^{-17} $ G on Mpc scales  \cite{ref3,ref4}.
Magnetic fields are  also present in accretion disks  such as 
protoplanetary disks
and  play an 
important  role in disks evolution.      
They  induce disk accretion via
magneto-rotational instability \cite{Balbus-Hawley} and  also produce  disk winds 
and outflows via the 
magneto-centrifugal force \cite{Blandford-Payne} or the magnetic pressure
\cite{Shibata-Uchida}.
A key parameter in the  understanding of the evolution of accretion disks
is the large-scale poloidal magnetic field  threading the  disks \cite{Muto}.
Moreover, although for the general scenario the magnetic fields present in the galaxies 
are toroidal,  poloidal magnetic fields  have been also  observed 
in the central regions  of galaxies  which have gaseous rings \cite{Lesch}.
Finite size current disks  are also  used as models of 
the magnetic dipole field presents in stars \cite{Toropin}. 
Such fields are small enough not to  influence  the spacetime structure of 
these objects.  Indeed, the general relativistic effects  are important for very compact objects 
and strong magnetic fields 
as  in  the case of  highly magnetized neutron stars, or magnetars,
where the magnetic field is  characterized by a  strength of the order of  
$10^{14}$ G \cite{neu1, neu2, neu3}. Thus, the Newtonian approximation is valid
for modeling  above structures. 

A simple model of thin disk is the  Kuzmin-Toomre disk
\cite{KUZMIN,TOOMRE}   which  represents a disk-like   matter distribution  with
a concentration of mass in its center and density that
decays as $ 1/r^3$ on the plane of the disc.  These  models
are constructed using  the image method that is usually used to solve problems
in
electrostatics. This structure has no boundary of the mass
but   as   the surface mass  density
 decreases rapidly  one can define a cut off radius, of the order of
the  galactic disk radius,  and, in principle, to
consider these disks as finite.

Another simple set of models are the Morgan-Morgan disks \cite{Morgan1}. These
disks have a mass concentration on their centers and finite
radius.  The models are constructed using a method developed
by Hunter \cite{Hunter} based on obtaining solutions of Laplace
equation in terms of oblate spheroidal coordinates, which are
ideally suited to the study of flat disks of finite extension.
Several classes of analytical models 
corresponding to   thin disks
have been obtained by different authors % [\ref{bib:Bell} - \ref{bib:Rein}].
\cite{Bell, Kalnajs, Conway, Gon-Reina, Pe-Ra-Gon, Gon-Pla-Ra, Ra-Ag-Pe, Rein}.
%Bell, Kalnajs, Conway, Gon-Reina, Pe-Ra-Gon, Gon-Pla-Ra, Ra-Ag-Pe, Rein
Thin disks  with electromagnetic fields, specially in curved spacetimes,  also
have been studied extensively  %[\ref{bib:KBL} - \ref{bib:Gut}].
\cite{KBL, LET1, GG1, Guti1, GG-CRIS, Guti2,AN-GUI-QUE, AN-GON-GUI,Gut}.

In this work we consider  analytical  models of finite thin disk
in presence of a  magnetic field. The mass density is determined by  Morgan-Morgan
solutions  and  the source   of magnetic field is constructed using  the Hunter
method  which involves the use of oblate spheroidal coordinates.
It produces as source an  azimuthal (toroidal) electric current distribution
which extends on all the plane of the disk, and hence a poloidal  magnetic field. 
These models differ from those presented in foregoing  references in that
we assume that  magnetic field strength  is known  far away from the disk  
and then  we determine the magnetic field structure near to the  disk.
The radial  component of the  magnetic field  on the disk surface 
or the profile of the azimuthal  electric current density are also assumed.

The paper is organized as follows.  In Section  2  we present
the general formalism  to construct    models of  thin disks  with  electric current.
We also analyse  the motion of  charged  test particles   around of the disks
and  we derivate
the stability condition of the system   against radial perturbations
using an extension  of the Rayleigh criterion.

In Section 3, we first present a summary
of the pair potential-density associated
to the Morgan- Morgan disks. 
Then  we construct  a family of finite  thin  disk with  electric  current in presence of a pure
magnetic field. As a simple example,  we consider  a particular profile 
of  azimuthal  electric current density with a physically reasonable behavior  
and we present explicit
expression
for the main  physical quantities associated to the disks  for the two first
terms of the series of the solutions. 
Finally, in Section  4   we summarize  the results obtained.

% un perfil de correiente especifico    .....   

%---------------------------------------------------------------------
%---------------------------------------------------------------------

%\end{document}

\section*{2. Disks and Maxwell equations  }

Exact solutions of Poisson's  equation representing the field of a thin disk
at $z = 0$  can be constructed assuming the gravitational potential $\Phi$ 
continuous 
across the disk, and its first derivative discontinuous in the direction normal
to the disk.
This can be written as
\begin{equation}
[\Phi_{,z}] =  2   \left . \Phi_{,z}     \right|_ {z=0^+}.
\end{equation}
%\begin{subequations}\begin{eqnarray}
 %& [\Phi_{,z}] = & 2   \left . \Phi_{,z}     \right|_ {z=0^+} ,  \\
 %&[ A_{,z}]=& 2  \left .  A_{,z}   \right|_ {z=0^+}  ,
%\end{eqnarray}\end{subequations}
Since the disk is thin the Poisson's equation  can be written as
\begin{equation}
\nabla ^2 \Phi = 4 \pi G \Sigma (R) \delta (z) , \label{eq:poisson}
\end{equation}
where $ \delta (z) $ is is the usual Dirac function with support on the disk
and $\Sigma(R)$ is the surface mass density.
%which for axially symmetric  fields
%is a function of cylindrical coordiantes  $R$ and $z$ only.
The mass density can be obtained,
for example, by  using the approach presented in Ref. \cite{CHGS}.
Thus, written by  the Laplacian operator in cylindrical coordinates and
integrating
from  $z=0^-$ to   $z=0^+$, we obtain
\begin{equation}
\Sigma (R)=  \frac{1}{2 \pi G}   \left .    \frac{ \partial \Phi }{\partial z}  
\right|_ {z=0^+} .  \label{eq:sigma}
\end{equation}

Similarly, thin  disks with  electric  current in presence of a pure magnetic
field  can be
obtained   assuming the magnetic potential ${\bf A}$  continuous and its first
derivate    discontinuous. That is
\begin{equation}
[ {\bf A}_{,z}]= 2  \left .  {\bf A}_{,z}   \right|_ {z=0^+}.
\end{equation}

The  magnetic field is governed by  the Maxwell's equations
\begin{subequations}\begin{eqnarray}
\nabla \cdot {\bf B} & =& 0 ,  \\
\nabla  \times {\bf B} & =& \mu _0 J \label{eq:max} ,
\end{eqnarray}\end{subequations}
where $J$ is the electric current density vector.
For axially symmetric  fields the magnetic potential is  ${\bf A} = A  \hat {\bf
\varphi}$
where $A$ is function of $R$ and $z$ only and $ \hat {\bf \varphi}$ a unit
vector in the azimuthal direction. Since
\begin{equation}
{\bf B} = \nabla \times {\bf A},  \label{eq:BA}
\end{equation}
 Eq. (\ref{eq:max}) can be written  as
\begin{equation}
 \nabla^2 A - \frac{1}{R^2} A = - \mu _0  {\text j}_\varphi  \delta (z) 
\label{eq:A}.
\end{equation}
where ${\text j}_\varphi$
is the surface  azimuthal electric current density.  It implies that the magnetic field is poloidal. 
Again written by  the Laplacian operator in cylindrical coordinates and
integrating  from  $z=0^-$ to   $z=0^+$, we obtain
\begin{equation}
\text{j}_\varphi = - \frac{2}{ \mu_0}   \left . \frac{ \partial A }{\partial z} 
 \right|_ {z=0^+} .  \label{eq:j}
\end{equation}

%rapidez de las particulas

Now we  analyse   the motion of  charged  test particles   around of the disks.
For equatorial circular orbits  the equations of motion of  test particles read
\begin{equation}
 \frac{\partial \Phi}{\partial R} -  \tilde e v \frac {1}{R} \frac{\partial (R
A)}{\partial R} = \frac {v^2} {R},  \label{eq:motion}
\end{equation}
where $\tilde e$ is the specific electric  charge of the particles and
 $v$ is the speed of the particles which is given by
\begin{equation}
 v_c = - \frac {\tilde e}{2}  \frac{\partial }{\partial R} (R A) \pm \sqrt { 
\left (\frac {\tilde e}{2}
 \frac{\partial }{\partial R} (R A)    \right )^2 + R  \frac{\partial \Phi 
}{\partial R} }.
 \label{eq:vc}
\end{equation}
The  positive sign corresponds to the direct orbits or co-rotating
and the negative sign  to the retrograde orbits or counter-rotating.

%Condicion de estabilidad

To analyse the stability of  the particles of the disks  in the case of
circular orbits in the equatorial plane we
use  an extension of Rayleigh  criteria of stability of a fluid at rest
in a gravitational field 
% [\ref{bib:RAYL} - \ref{bib:LETSTAB}].
\cite{RAYL,FLU,LETSTAB}.
The  method works as follows.   Any small element
of the matter distribution  analyzed  (in our case  a test particle in the
disk) is  displaced  slightly   from its path.
As a result of this displacement, forces appear which
act on the displaced matter  element.  If the matter distribution is stable,
these forces must tend to return the element
to its original position.

The first term  on the left-hand side  of the motion Eq. (\ref{eq:motion}) 
is
the gravitational force
$F_g$ (per unit of mass), the second term  the magnetic  force $F_m$,  and the
term on the
right-hand side  the centrifugal force $F_c$  acting
on the test particle.
So  we have a balance between the total force $F =F_g +F_L$
and the centrifugal force.   We
consider the particle to be initially in a circular orbit with
radius $R=R_0$.  In terms of specific angular momentum $L = Rv + \tilde e AR$,
$F_c(R_0)= (L_0-\tilde e A R_0) ^2/R_0^3$.  We slightly displace  the particles 
to a higher orbit
$R>R_0$. Since the angular momentum of particle  remains equal to its initial
value, the centrifugal force   in its new
position is $\hat  F_c(R) = (L_0-\tilde e A R) ^2/R^3 $. In order that the
particle returns to its
initial position must be met that  $F(R)>\hat F_c(R)$, but
according to the balance Eq.  (\ref{eq:motion})
$F(R)=F_c(R)$ so that $F_c(R) >\hat  F_c(R)$,  and  hence
$(L -  \tilde e AR)^2 > (L_{0} -  \tilde e A R)^2 $.
It follows that  the condition of stability is $L^2 > L_0^2 $.   By doing a
Taylor expansion of
$L$ around $R=R_0$, we can write this condition in the form
\begin{equation}
L  \frac{dL}{dR}  >0,
\end{equation}
or, in other words, $d L^2 / dR >0$.

From Eq.  (\ref{eq:BA})   we find
\begin{equation}
 \frac{\partial }{\partial R} (R A) dR + \frac{\partial }{\partial z} (R A) dz
=0.
\end{equation}

Thus the Eq. $R A = C $, with $C$ constant,  gives  the lines
of force of the magnetic field.

%---------------------------------------------------------------------
%---------------------------------------------------------------------

\section*{3. Finite disks with magnetic field }

\subsection*{3.1. Morgan-Morgan Disks}

Solutions representing the field of a finite thin disk can be obtained  solving
the Laplace equation in oblate  spheroidal coordinates
($u$,$v$), which are defined in terms of the cylindrical
coordinates  ($R$, $z$)  by
%$x=-iu$,  $y=v$, and  $k=i a$
\begin{subequations}\begin{eqnarray}
 R^2 & =& a ^2 (1 + u^2 )(1 - v^2 ) ,  \\
 z & = & auv,
\end{eqnarray}\end{subequations}
and explicitly
\begin{subequations}\begin{eqnarray}
 \sqrt 2 u  & = &  \sqrt { [ ( \tilde R^2 + \tilde z^2 -1)^2 + 4\tilde z ^2
]^{1/2} +  \tilde R^2 + \tilde z^2 -1 } ,  \\
 \sqrt 2 v  & = & \sqrt { [ ( \tilde R^2 + \tilde z^2 -1)^2 + 4\tilde z ^2
]^{1/2} - ( \tilde R^2 + \tilde z^2 -1) },
\end{eqnarray}\end{subequations}
where $u \geq 0$ , $-1< v <1$, $\tilde R = R/a$, $\tilde z = z/a$, and  $a$ the
radius of the
disk.  The disk is located in  $u=0$,  $-1< v <1$,
and   when it  is crossed the coordinate $v$ changes of sign but not its
 absolute value, whereas $u$ is continuous.
This implies that an even function of $v$ is a continuous function
everywhere but has a discontinuous $v$ derivative at the
disk.

In this coordinate system the Laplacian operator  has the form
\begin{equation}
\nabla^2 = \frac{1}{a^2(u^2 + v^2)} \left [
\frac{\partial}{\partial u} (1 + u^2) \frac{\partial}{\partial u}  +
\frac{\partial}{\partial v} (1 - v^2) \frac{\partial}{\partial v} \right ]. 
\label{eq:laplaciano}
\end{equation}

The   general solution of Laplace’s equation  can be written as
\begin{equation}
 \Phi  = - \sum\limits_{n = 0}^\infty  {c_{2n} q_{2n} (u)P_{2n} (v)} ,
\label{EQ:serieobla}
\end{equation}
where  $c_{2n}$ are constants, $P_{2n}$ are the Legendre polynomials of order
$2n$ and
\begin{equation}
q_{2n} (u) = i^{2n + 1} Q_{2n} (iu),
\end{equation}
being  $Q_{2n} (iu)$  the Legendre functions of the second kind.

The mass surface density  Eq.  (\ref{eq:sigma}) takes the form
\begin{equation}
\Sigma(R) = \frac{1}{2 \pi a G v} \sum\limits_{n = 0}^\infty  c_{2n} (2n+1)
q_{2n+1} (0)P_{2l} (v)
\end{equation}
with $v= \sqrt{1-R^2/a^2}$.

For $n=0$ we have  the zeroth order Morgan-Morgan
disk  and,
 for the two first terms of the series,
 the first Morgan-Morgan disk \cite{LETSTAB}.
In this case,  the gravitational potential    is
\begin{equation}
\Phi   = -   \frac{MG}{a} \left \{ \cot^{ - 1} (u) + \frac{1}{4}
 \left[ (3u^2  + 1) \cot^{ - 1} (u) - 3u \right] \left( 3v^2  - 1 \right) \right
\},
\end{equation}
being  $M$ is the mass  of the disks, and the surface mass density
\begin{equation}
\Sigma(R) = \frac{3M}{2 \pi a^2} \sqrt{ 1-\frac{R^2}{a^2} }. \label{eq:sig1}
\end{equation}

\subsection*{3.2. Current  Disks}

 We now  consider a thin  disk with  electric  current in presence of a pure
magnetic field.
Using  the  Laplace operator in the form of Eq. (\ref{eq:laplaciano}),  making the change of variable $u=ix$, using  the algebraic relation
\begin{equation}
\frac{x^2-v^2}{(1-x^2)(1-v^2)} = \frac{1}{1-x^2} -  \frac{1}{1-v^2},
\end{equation}
and letting $A = X(x) V(v)$,  Eq. (\ref{eq:A}) for the magnetic potential   yields two associated Legendre equations of first order
\begin{subequations}
\begin{eqnarray}
(1-x^2) \frac{d^2 X}{dx^2} -2x \frac{d X}{dx} + l(l+1)X  - \frac{1}{1-x^2}X & =
& 0,  \\
(1-v^2) \frac{d^2 V}{dv^2} -2v \frac{d V}{dv} + l(l+1)V  - \frac{1}{1-v^2}V & =
& 0.
\end{eqnarray}
\end{subequations}
The solution which vanishes at infinity is  $A_n = b_n Q^1_{n} (-iu) P^1_{n}(v) 
$,
where $b_n$ is a constant, $P^1_n(v)$ are the associated  Legendre functions
of the the first kind of order $1$ and
$Q^1_{n} (-iu)$  the associated Legendre functions of the second kind of order
$1$ \cite{Abra, Morse}.
But
$Q^m_n (-z) = (-1)^{ n+m + 1} Q^m_n (z)$, hence  $A_n = b_n (-1)^{n+2} Q^1_{n}
(iu) P^1_{n}(v)$.
Moreover, since A must be invariant to reflection in the equatorial plane by
symmetry of the problem, $A(u,v)=A(u,-v)$,
the parity property $P^m_n (-v) = (-1)^{  n + m } P^m_n (v)$  \cite{Arfken}
shows that
$n$ is an odd integer. Thus the most general solutions for the  magnetic
potential  can be written as
\begin{equation}
 A  =  \sum\limits_{n = 0}^\infty  {b_{2n+1}  (-1)^{2n+3} Q^1_{2n+1} (iu)
P^1_{2n+1} (v)} .
 \label{eq:serieA}
\end{equation}

As $A$ is a even  function of $v$, it is  continuous across the disk. 
Consequently,   its
normal derivate is an  odd  function of $v$ and  hence discontinuous across the
disks.
This implies that   the source of the magnetic field also is  planar.
Using the relation
\begin{equation}
 \frac{d}{dz} Q^m_n (z)  \mid_{z=0} = - (n-m+1) Q^m_{n +1}(0),
\end{equation}
one  finds that the surface current density  Eq. (\ref{eq:j}) is
\begin{equation}
\text{j}_ \varphi = \frac{2}{\mu_0 a  v} \sum\limits_{n = 0}^\infty   b_{2n+1}
(2n+1)(-1)^{2n+3}
i  Q^1_{2n+2} (0) P^1_{2n+1} (v),
\end{equation}
where $v=\sqrt{1-R^2/a^2}$.

% Calculo de los coeficientes b_{2n +1}

In order to determine the coefficients  $b_{2n+1}$, we   write the current density
as
\begin{equation}
 \text{j}_ \varphi =\frac{2}{\mu_0 a  } \frac{ F(v)}{v}
\end{equation} 
with  
\begin{equation}
 F(v) = \sum\limits_{n = 0}^\infty  a_{2n +1}  P^1_{2n+1} (v),
\end{equation}
where $a_{2n +1} = b_{2n+1}(2n+1)(-1)^{2n+3} i  Q^1_{2n+2} (0) $.
 By using the orthogonality
property of the  the associated  Legendre functions
of the the first kind, we obtain
\begin{equation}
 a_{2n +1} = \frac{(4n + 3) (2n)!}{2 (2n +2)! } \int _{-1}^{1} F(v) P^1_{2n+1}  dv \label{eq:coef1}.
 \end{equation}

On the other hand,   from  Eq. (\ref{eq:BA}) we have 
 \begin{equation}
 B_R = - \frac { \partial A } {\partial z} ,  \label{eq:BR}
 \end{equation}
 %B_z= (1/R) \frac { \partial} {\partial R ]  (AR)
so that Eq.  (\ref{eq:j}) for  the current density  can be written as 
  \begin{equation}
\text{j}_\varphi = - \frac{2}{ \mu_0}     B_R |_ {z=0^+} .  \label{eq:hola}
\end{equation}
 
 Thus, using the identity
\begin{equation}
iQ^1_{2n+2} (0) = (-1)^{n+2}(2)^{n+1}  \frac{ (n+1)! }{ (2n+1)!! },
\end{equation}
 we get 
 \begin{equation}
 b_{2n +1} = \frac{a (4n + 3) [(2n)!]^2}{ (-1)^{3n} (2)^{2n+2} n! (n+1)! 
 (2n +2)! } \int _{-1}^{1} v B_R (v) P^1_{2n+1} (v)  dv \label{eq:coef1}.
 \end{equation}
This expression permits us  to determinate all  the coefficients  $b_{2n +1}$  
by giving  the radial component  of the magnetic field on the disk
or the profile of the azimuthal  electric current density.   

%An expression determining all b_{2n+1} for given imposed magnetic field would surely 

\subsection*{3.3. A simple example }

As a simple example, we consider  the azimuthal current density of the  
disk  to be
 \begin{equation} 
 \text{j}_ \varphi = j_0 \left ( \frac Ra \right )  \left ( 1- \frac {R^2}{a^2} \right ) ^{ \frac{j_1}{2} }, 
 \end{equation}
 where $j_0$ and $j_1$ are constants.
 In oblate coordinates we have 
 \begin{equation}
 \text{j}_ \varphi = j_0 v^{j_1} (1-v^2)^{ \frac{1}{2} }.
 \end{equation}
%Now, it is clear that the surface density
%diverges at the disc edge, when η = 0, unless that we impose the
%condition (Hunter 1963)
%Let us assume
%Outside of the thin disc, the azimuthal current density vanishes
Since  the surface current  density
diverges at the disc edge, when $v = 0$,
this expression satisfies the  requirement that 
the current density must vanish  there.
It also vanishes  at the disk center, which can indicate 
the presence of a central star. 

By taking  $j_1$ as a   odd positive  integer  the integral in Eq.  (\ref{eq:coef1}) is even 
and hence 
\begin{equation}
 b_{2n +1} = \frac{\mu_0 a j_0  (4n + 3) [(2n)!]^2}{ (-1)^{3n+1} (2)^{2n+2} n! (n+1)! 
 (2n +2)! } \int _{0}^{1} v^{j_1+1} (1-v^2)^{ \frac{1}{2} } 
 P^1_{2n+1}  dv.
 \end{equation}

Using the expression \cite{Bateman}
\begin{eqnarray}
(-1)^m 2^{m+1} \Gamma (1-m + n) \int _0 ^1 x^\sigma (1-x^2)^{\frac m 2}
P_n ^m (x) dx  =&   \nonumber  \\  %\right .   \nonumber  \\  \left .
  \frac{\Gamma (\frac 12 +  \frac 12 \sigma)  \Gamma (1 + \frac 12 \sigma) 
\Gamma (1 + m + n)}
{\Gamma (1 + \frac 12 \sigma +\frac 12 m - \frac 12 n )  \Gamma (\frac 32 +
 \frac 12 \sigma + \frac 12 m +\frac 12 n )}, &
\end{eqnarray}
which is valid for $\text{Re} \  \sigma > -1$ and $m$ a positive integer,  
we  obtain
%\begin{equation}
 %b_{2n +1} = \frac{\mu_0 a j_0(4n + 3) (2n)!}{8 (-1)^{2n+4} (2n +2)!
%(2n+1) i  Q^1_{2n+2} (0) } 
%\frac{ \Gamma ( 1 +  \frac 12 j_1)  \Gamma (\frac 32 + \frac 12 j_1) 
%\Gamma (3 + 2n)}
%{ \Gamma ( 2n +1) \Gamma (\frac 32 + \frac 12 j_1 -n )  \Gamma ( 3 +
% \frac 12 j_1 + n )},
 %\end{equation}
\begin{eqnarray}
 b_{2n +1} &=& \frac{\mu_0 a j_0(4n + 3) [(2n)!]^2}{ (-1)^{3n} 2^{2n+4} n! (n+1)! (2n +2)! }  \nonumber   \\
&  & \times \frac{\Gamma (2n + 3)  \Gamma ( 1 +  \frac 12 j_1)  \Gamma (\frac 32 + \frac 12 j_1) 
}
{ \Gamma ( 2n +1) \Gamma (\frac 32 + \frac 12 j_1 -n )  \Gamma ( 3 +
 \frac 12 j_1 + n )},
 \end{eqnarray} 
where  $n \geq 0$ and $j_1$ is an  odd positive  integer.
The constant $j_0$ can  be determined  from a reference magnetic field $B_0$, 
and accordingly is the parameter that controls the magnetic field.  Since 
the field  is poloidal,   the exterior field  to the disk asymptotically approaches
at infinity a uniform vertical field which  is also measurable \cite{Troland, Crutcher},  and hence
this value of magnetic field can be taken as reference.   Thus,   from   Eq. (\ref{eq:BA}) we have that
\begin{equation}
j_0 = \frac{  R_0 B_0 } { \left .  \frac { \partial (\tilde AR)} {\partial R }  \right|_ {R_0} },
\end{equation}
where $R_0$ is  the radial distance at which the imposed magnetic field is known, and 
$\tilde A =A/j_0$.  For protoplanetary disks an accepted value  for this field is 
$B_0 =10 \mu \text{G}$ for $R_0 = 100 $ AU  \cite{Muto}.

%CASO PARTICULAR

We   consider  the two first terms of the series in Eq. (\ref{eq:serieA}) taking $j_1 =1$. 
For  $n=0$  have $b_1= \frac {1}{20} \mu_0 a j_0$ and  for 
 $n=1$   $b_3=- \frac {1}{120} \mu_
0 a j_0$. Thus $b_1 = -6 b_3$ and the parameter $j_0$ takes the value 
\begin{equation}
j_0 = \frac{2B_0}{5 \tilde b_1 [-3u + (3\tilde R^2_0-2) \cot^{-1} (u)] },
\end{equation}
where $\tilde b_1 = b_ 1/j_0$,  $u=\sqrt{\tilde R^2_0-1}$ and $\tilde  R_0 =R_0/a$.

In this case   the magnetic potential is given by
\begin{eqnarray}
 A &=& \frac {b_1}{8} \sqrt{1-v^2}  \left \{  \frac{u [8 +  (13 +15 u^2)(5v^2 -1 )
] }{\sqrt{1+u^2 }}  \right . \nonumber \\
&& \left . - [8+  3(1+5u^2) (5v^2 -1 )] \sqrt{1+u^2} \cot^{-1} (u)   \right\},
\end{eqnarray}
and   surface  electric charge density is
\begin{equation}
\text{j}_ \varphi = -\frac{20 b_1}{\mu_o a} (R/a) \sqrt{ 1-\frac{R^2}{a^2} } .
\end{equation}

In order to  analyse  the motion of  charged  test particles   around of the disks,
we consider the first Morgan-Morgan disk.
In oblate coordinates,  Eq.  (\ref{eq:vc}) for the  circular speed   takes the form
\begin{equation}
v_c = - \frac {\tilde e}{2}  \left ( A + \frac{(v^2-1)}{v} \frac{\partial A
}{\partial v}  \right )
\pm \sqrt { \frac {\tilde e^2}{4}  \left ( A + \frac{(v^2-1)}{v} \frac{\partial
A }{\partial v}  \right )   ^2
 + \frac{(v^2-1)}{v} \frac{\partial \Phi }{\partial v} },
\end{equation}
so that inside the disk the circular speed of particles is given by
\begin{equation}
v_c = - \frac 5 8 \tilde e b\pi \tilde  R (3\tilde R^2 -2) \pm \sqrt{ \frac {25}{64} \tilde
e^2 b^2 \pi^2  \tilde  R^2  (3 \tilde R^2 -2)^2
 + 3 \pi \tilde  R^2} ,  % \label{eq:vc}
\end{equation}
whereas outside  the disk is
\begin{eqnarray}
 v_c  &=& - \frac 5 4 \tilde e b \tilde R [-3u + (3\tilde R^2-2) \cot^{-1} (u) ] 
  \pm \left (\frac {25}{16} \tilde e^2 b^2  \tilde R^2
 [-3u  \right .  \nonumber \\
 &&  \left . +  (3\tilde R^2-2) \cot^{-1} (u) ]^2 
 + \frac 3 4 [-2u + \pi \tilde R^2 -2\tilde R^2 \cot^{-1}
(u)] \right )^{1/2} , 
\end{eqnarray}
with $u=\sqrt{\tilde R^2-1}$.
% $v=0$

\begin{figure}
$$
\begin{array}{c}
%\text{j}_ \varphi  \\
\includegraphics[width=0.4\textwidth]{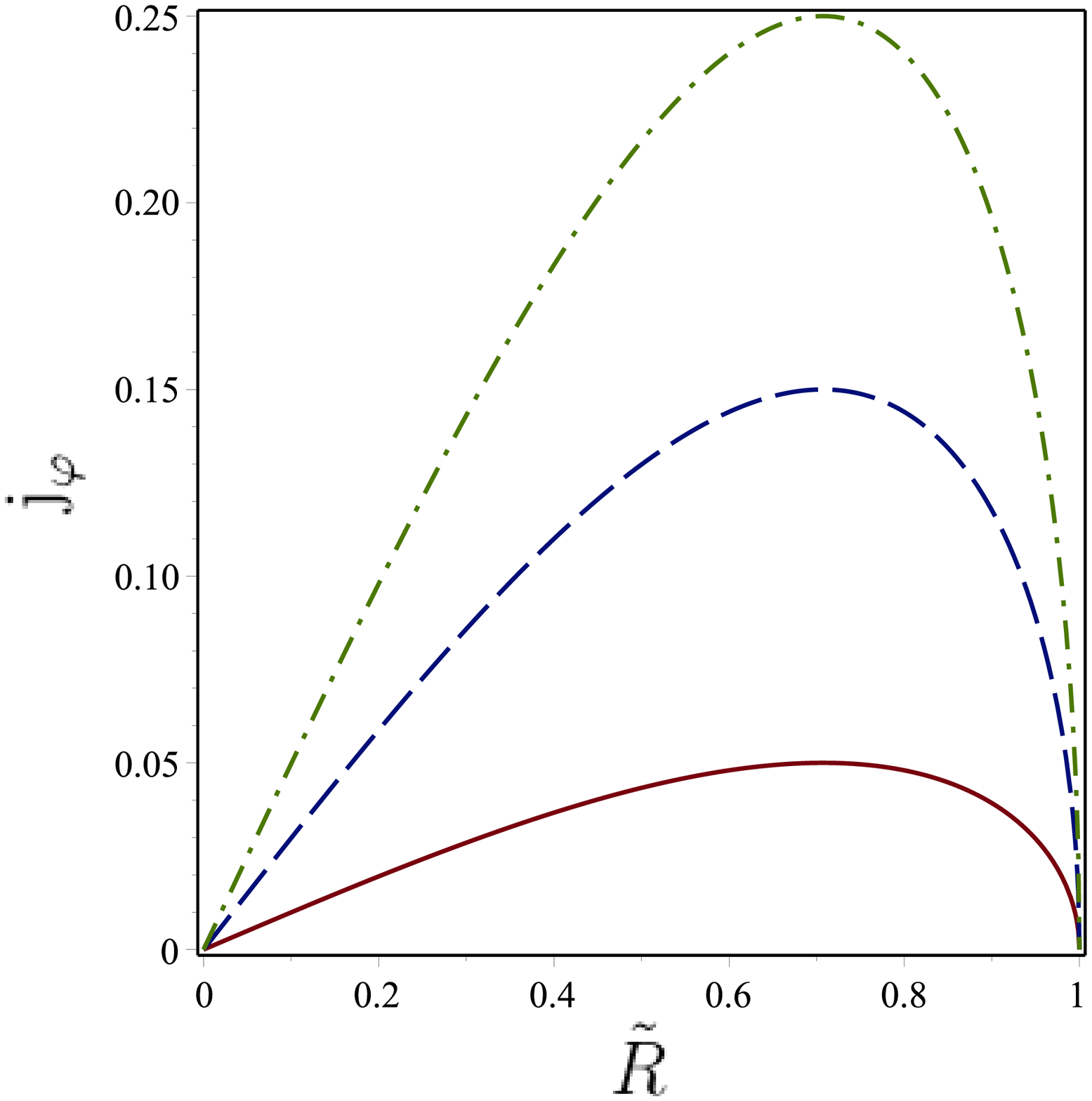}   \\
% \tilde  R   \\
\end{array}
$$	
\caption{ Electric current density $\text{j}_ \varphi$ for $\tilde e=1$  and
different values of
the parameter of magnetic fields $b_1= 0$ ($\tilde  R$ axis), $0.1$  (solid curve),  $0.3$ (dashed  curve), and 
$0.5$ (dash-dotted  curve),
as functions of $\tilde  R$. }
\label{fig:j}
\end{figure}

In Fig. \ref{fig:j}  we have plotted  the electric current density
$\text{j}_ \varphi$  for $\tilde e=1$ and   different values of
the parameter of magnetic fields  $b_1= 0$ , $0.1$ ,  $0.3$, and 
$0.5$,
as functions of $\tilde  R$.  We see that $\text{j}_ \varphi$ is zero in the center of
the disk
where the surface mass density is greater  (Eq. (\ref{eq:sig1})), reaches a
maximum and then
falls to zero at rim of the disk where there is no matter.

\begin{figure}
$$
\begin{array}{cc}
% v_+   & v_-  \\
\includegraphics[width=0.4\textwidth]{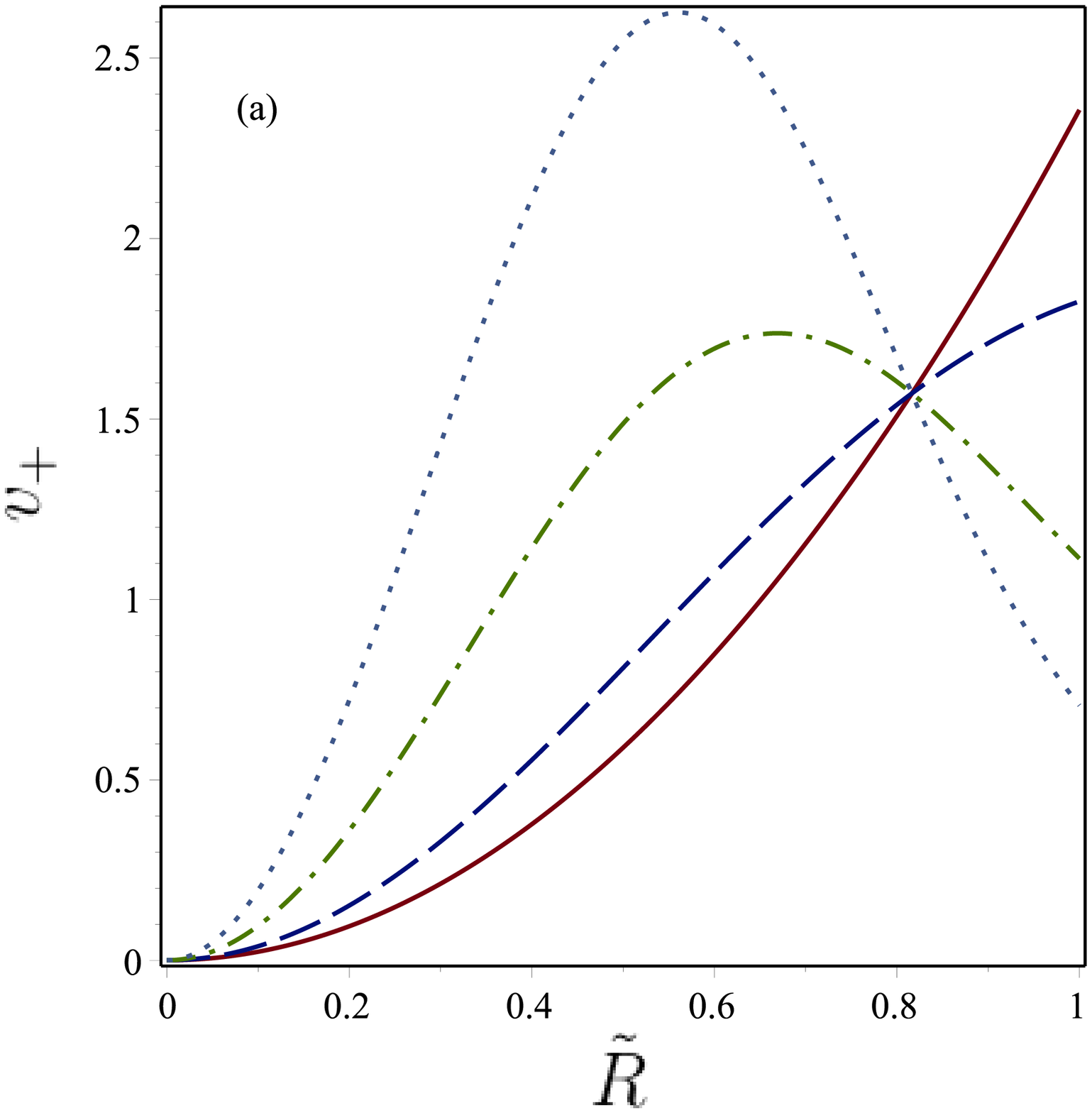} &
\includegraphics[width=0.4\textwidth]{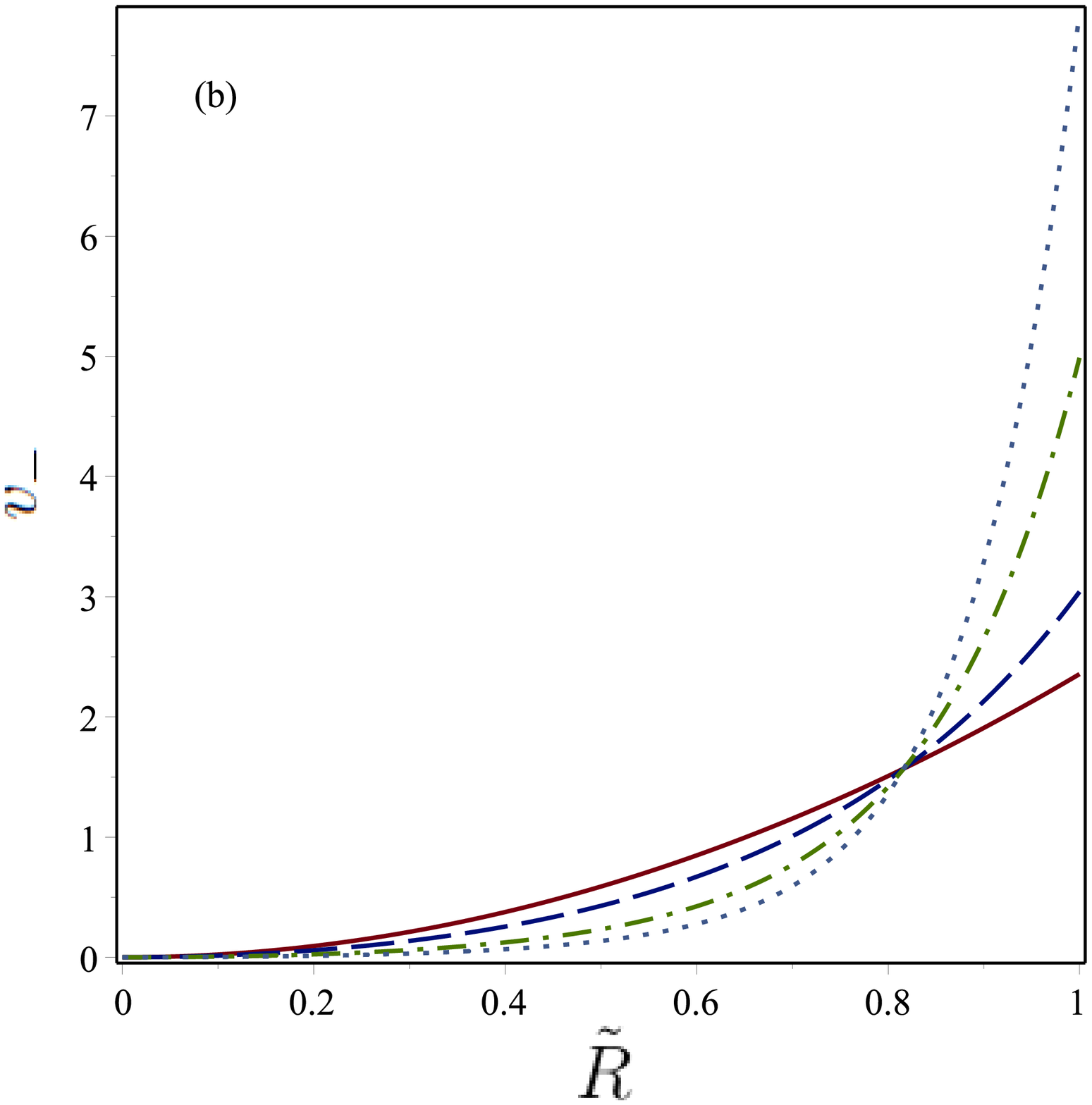}  \\
%\tilde  R & \tilde  R  \\
&  \\
% v_+   & v_-  \\
\includegraphics[width=0.4\textwidth]{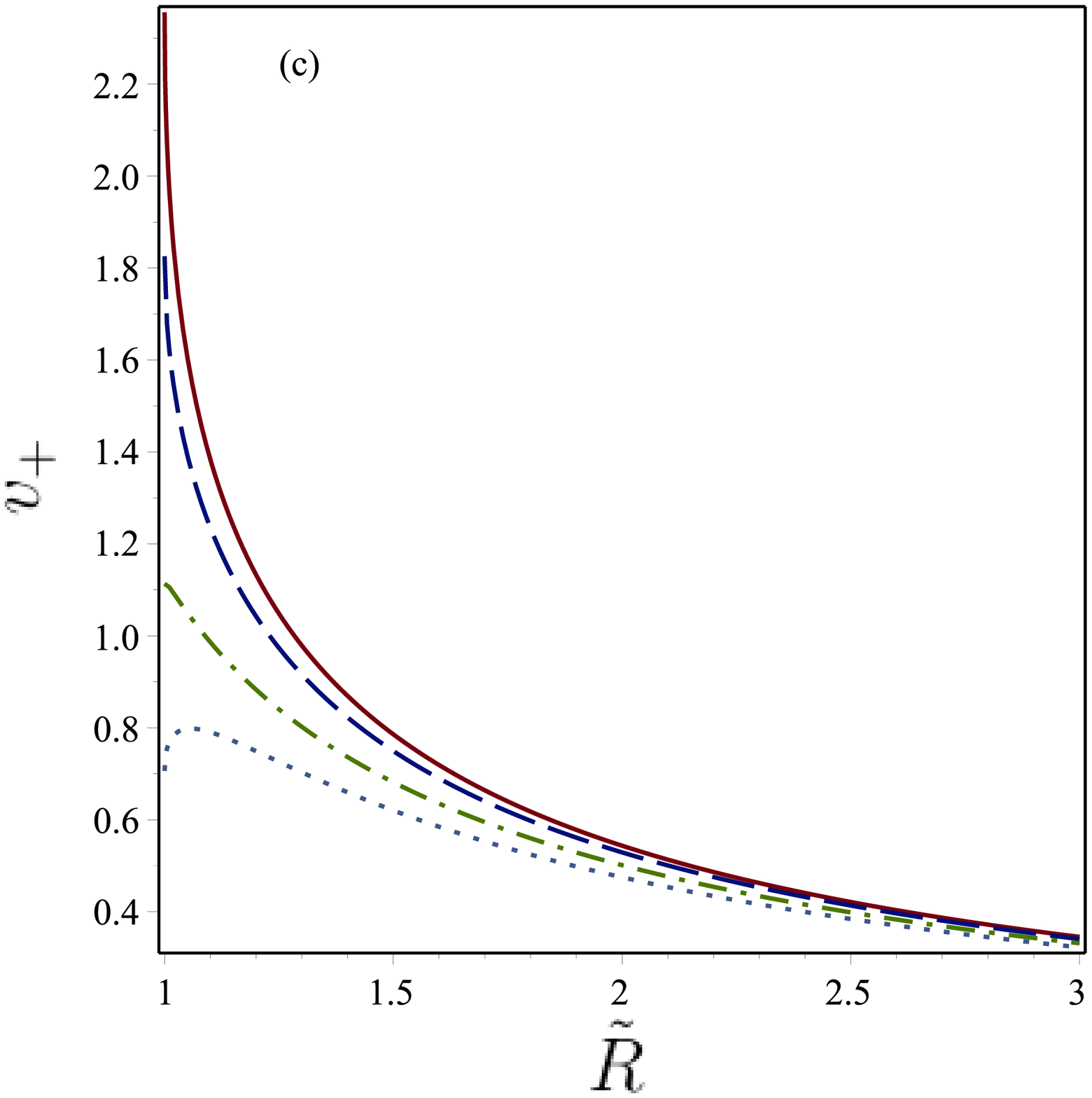} &
\includegraphics[width=0.4\textwidth]{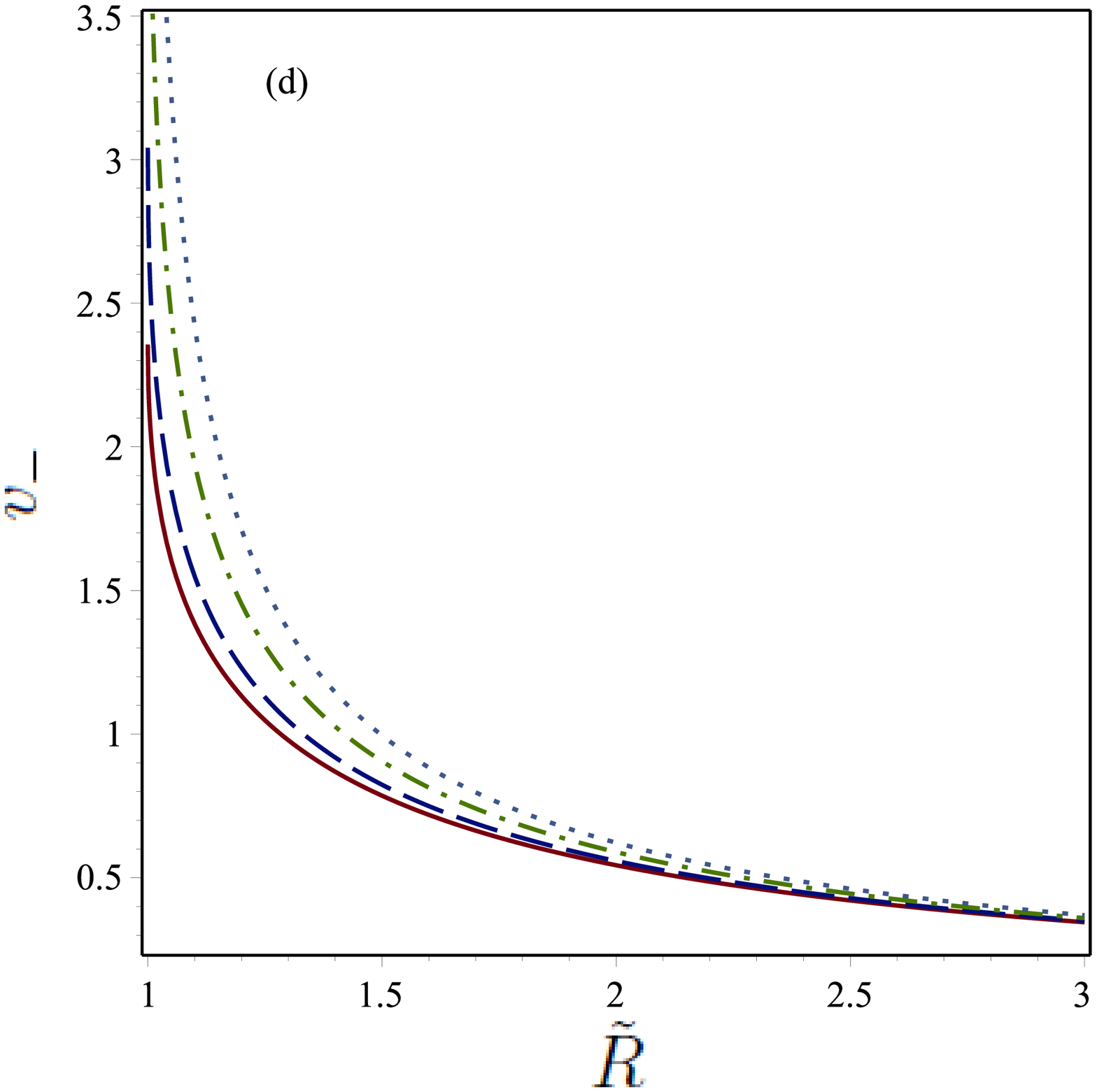}  \\
 %\tilde R & \tilde  R  
\end{array}
$$	
\caption{ $(a)$ Circular  speed  $v_+$ and  $(b)$  $v_-$  for the motion of test 
particles  inside of the the disk.
$(c)$ and  $(d)$ are the same 
for particles outside of the  disk   for $\tilde e=1$
and different values of
the parameter of magnetic fields $b_1= 0$ (solid curves), $0.1$ (dashed curves),  $0.3$ (dash-dotted curves ), and 
$0.5$ (dotted  curves),
as functions of $\tilde R$. }
\label{fig:v}
\end{figure}

In Fig. \ref{fig:v} we have  plotted   the curves of the  circular  speed  $v_+$ and  
$v_-$
for the motion of charged test  particles  inside  the the disk  (Figs.   \ref{fig:v}$(a)$
and   \ref{fig:v}$(b)$)
and for particles outside  the  disk (Figs.   \ref{fig:v}$(c)$ and   \ref{fig:v}$(d)$), for  $\tilde e=1$
and the same value of the
other parameters.
Inside the disk, we find  that  for direct orbits the magnetic field increases
the speed of particles
to a certain values of $\tilde R$, and then decreases it, whereas for retrograde orbits
the contrary occurs.
Outside the disk, we observe  that  for direct orbits the magnetic field 
decreases  the speed of particles
everywhere of the disk whereas for  retrograde orbits the contrary occurs.

\begin{figure}
$$
\begin{array}{cc}
% L^2_+   & L^2_-  \\
\includegraphics[width=0.4\textwidth]{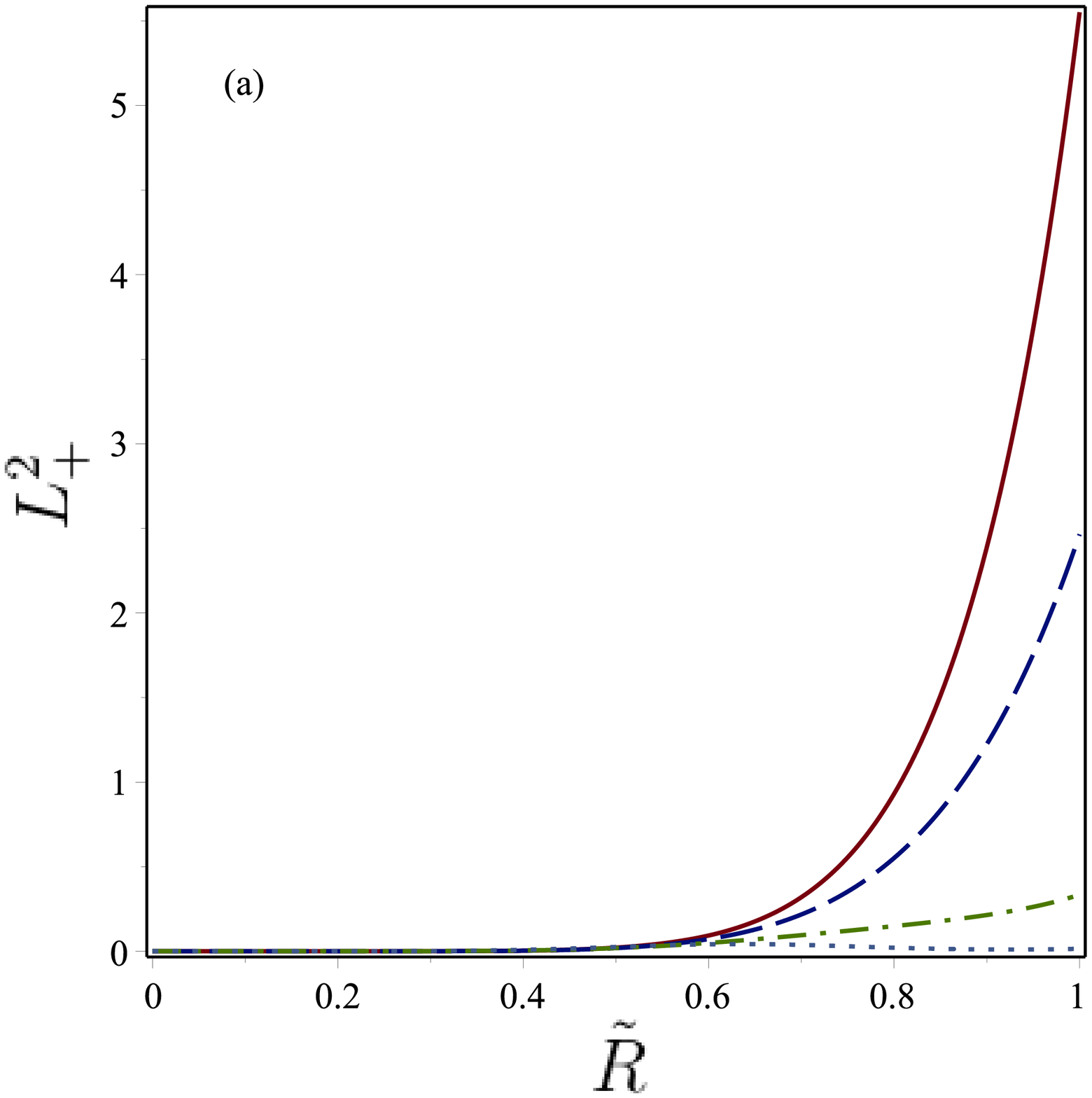} &
\includegraphics[width=0.4\textwidth]{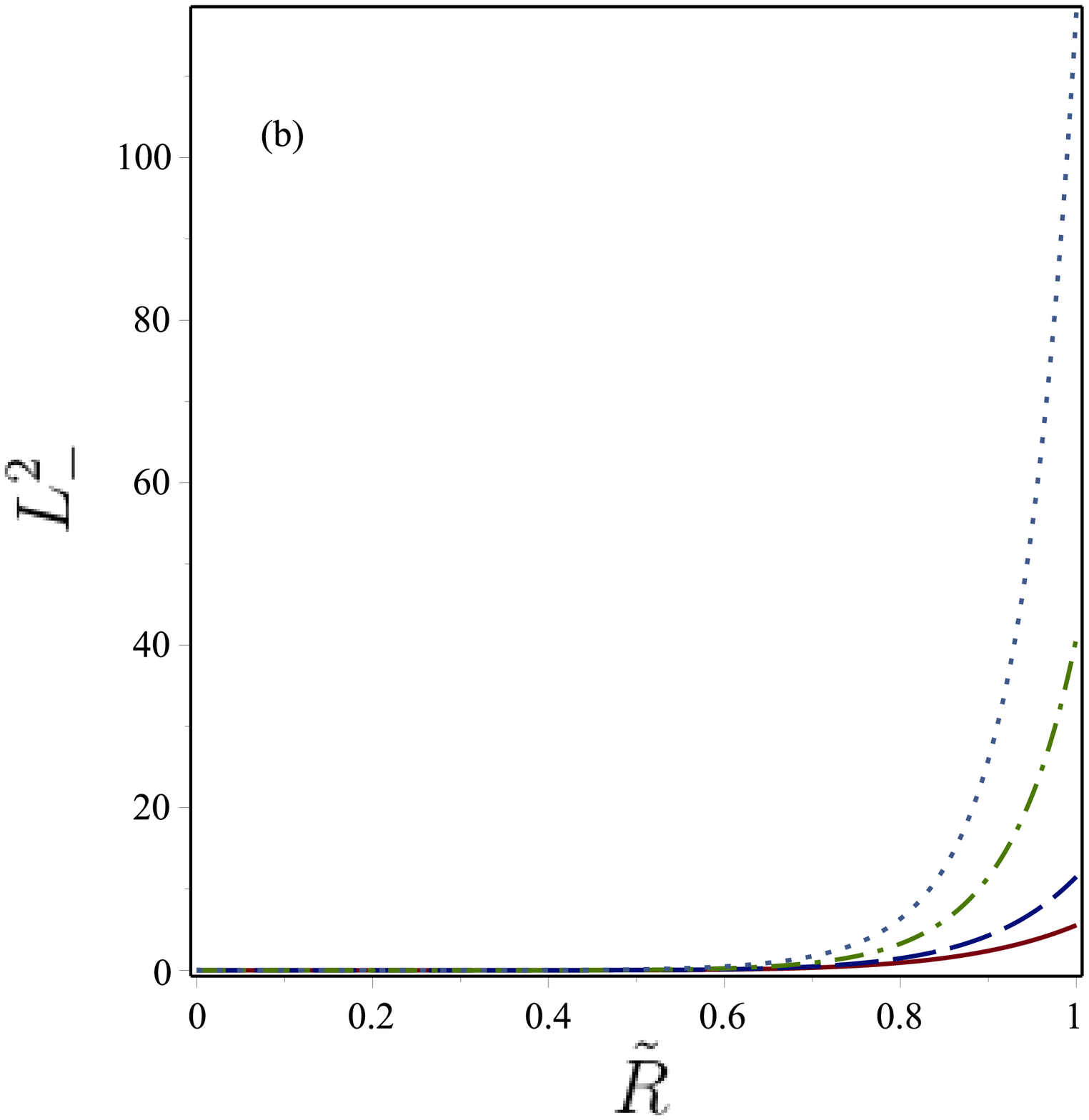}  \\
%\tilde  R & \tilde  R  \\
&  \\
% L^2_+   & L^2_-  \\
\includegraphics[width=0.4\textwidth]{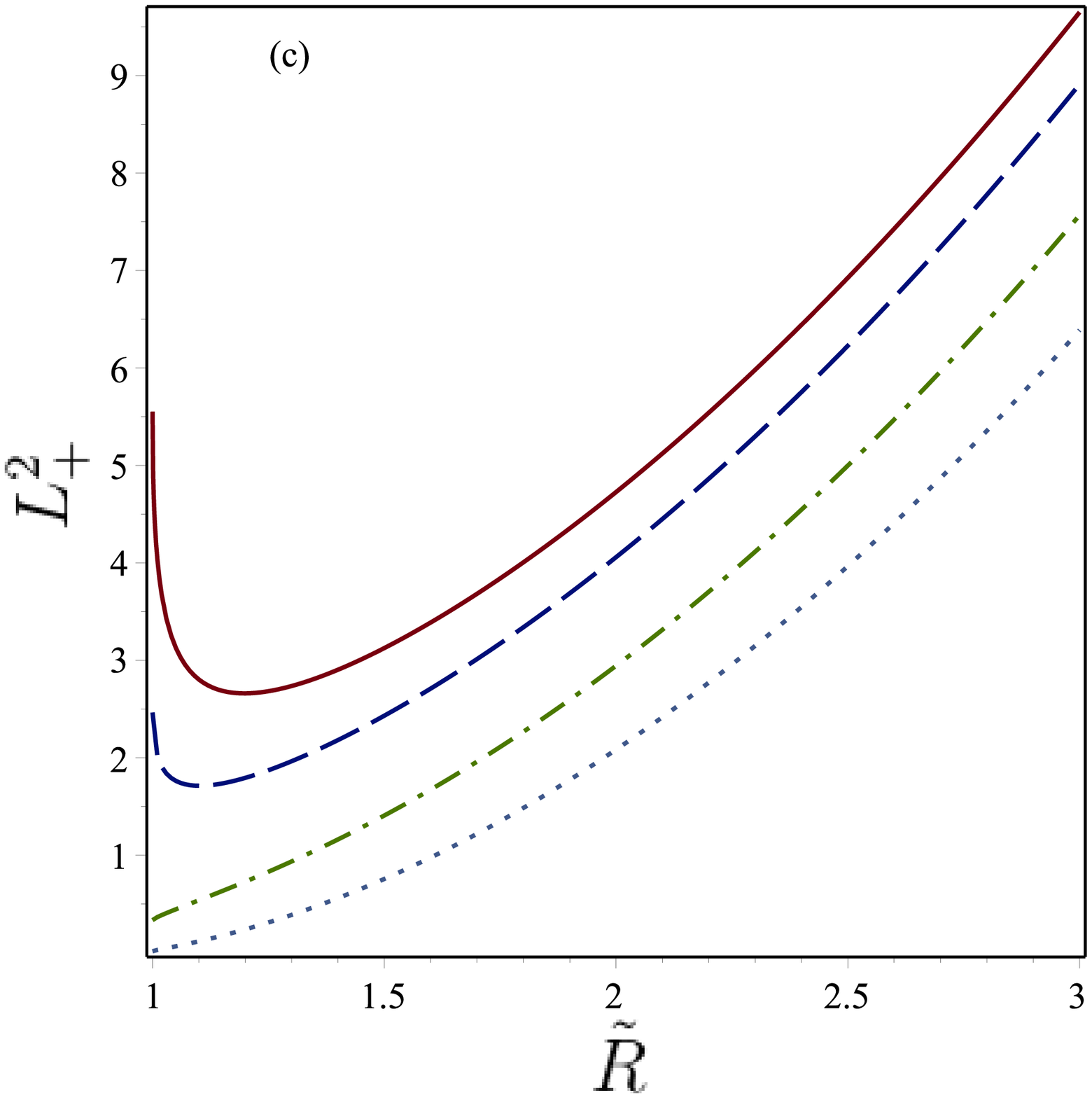} &
\includegraphics[width=0.4\textwidth]{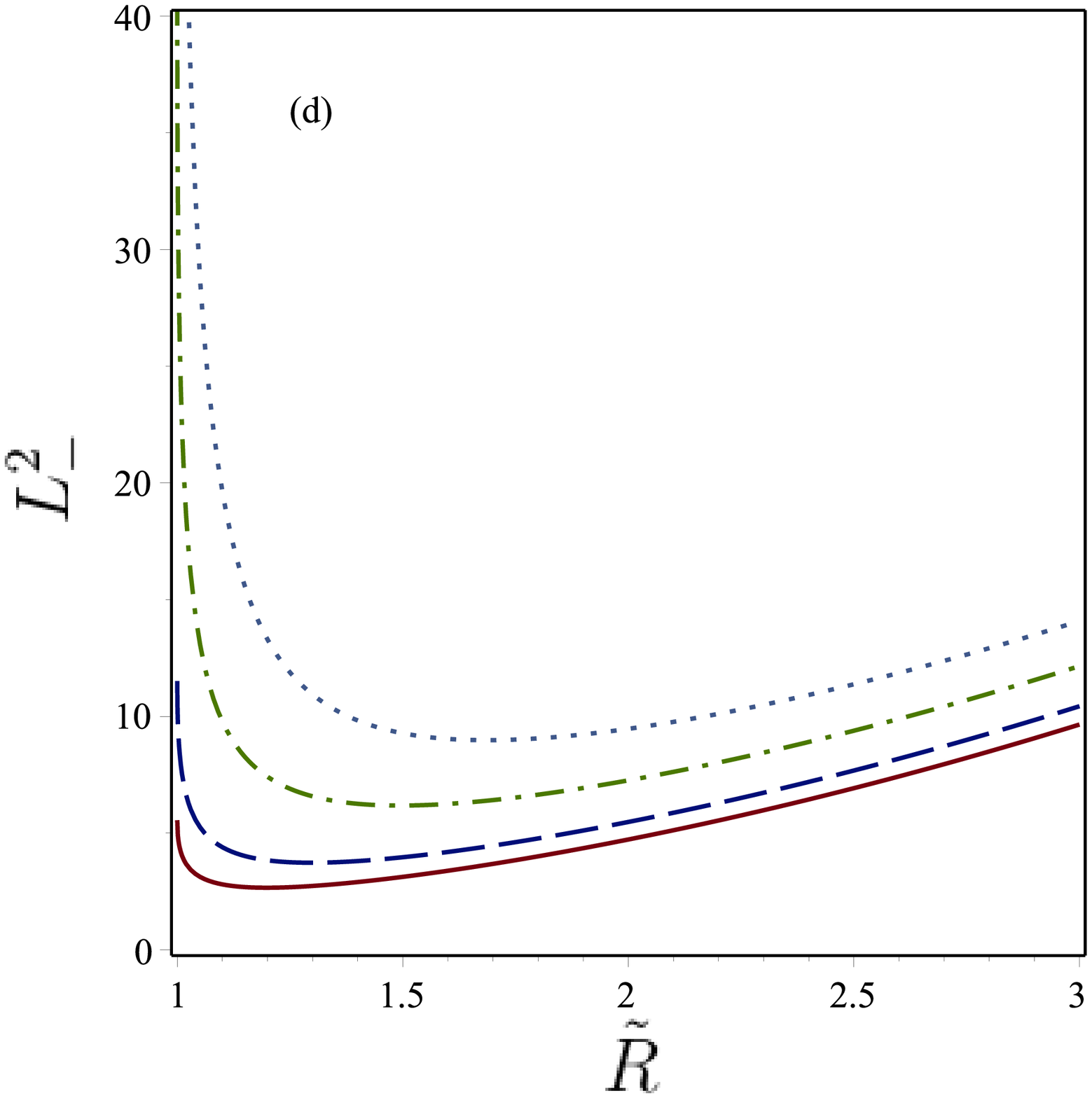}  \\
%\tilde  R &  \tilde  R 
\end{array}
$$	
\caption{$(a)$ Specific angular momentum    $L^2_+$ and   $(b)$  $L^2_-$  for the motion
of test  particles  inside of the the disk.
 $(c)$ and  $(d)$ are the same
for particles outside of the  disk  for $\tilde e=1$
and different values of
the parameter of magnetic fields $b_1= 0$ (solid curves), $0.1$ (dashed curves),  $0.3$ (dash-dotted curves ), and 
$0.5$ (dotted  curves),
as functions of $\tilde R$. }
\label{fig:L2}
\end{figure}

In Fig. \ref{fig:L2} we have  plotted   the specific angular momentum  $L^2_+$ and  
$L^2_-$
for the motion of test  particles  inside  the the disk  (Figs. \ref{fig:L2}$(a)$ and 
\ref{fig:L2}$(b)$)
and for particles outside  the  disk (Figs. \ref{fig:L2}$(c)$ and  \ref{fig:L2}$(d)$),  with $\tilde e=1$
and the same value of the
other parameters.
Inside the disk, we find that  for direct orbits the magnetic field can make
less stable
the motion of the particles against radial perturbations,
whereas for retrograde orbits the  particles are always stable.
Outside the disk, we observe  that  for direct orbits the magnetic field  can
stabilize   the particles,
whereas for retrograde orbits the magnetic field enhances the zone of
instability.

\begin{figure}
$$
\begin{array}{cc}

\includegraphics[width=0.4\textwidth]{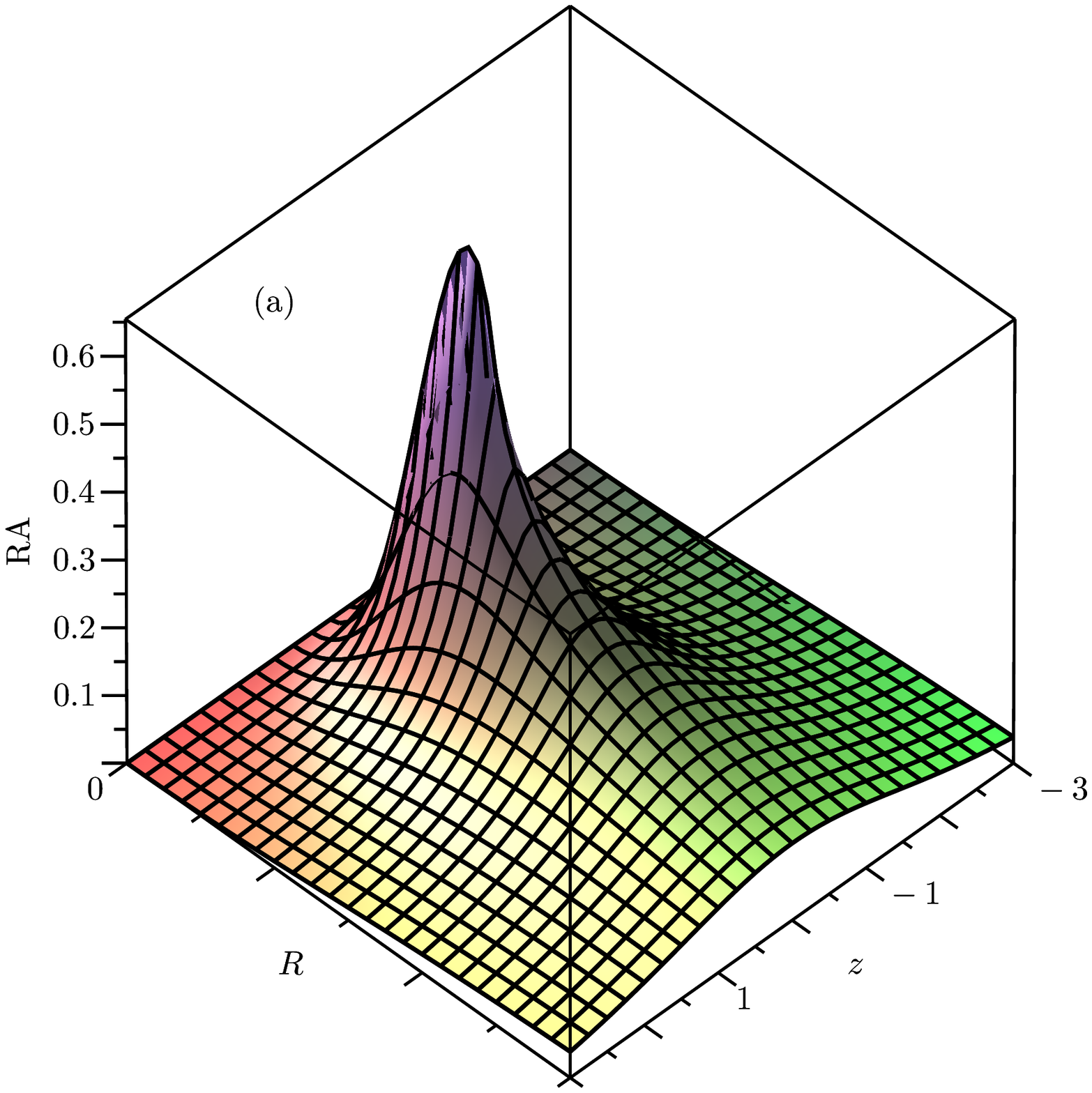} &
\includegraphics[width=0.4\textwidth]{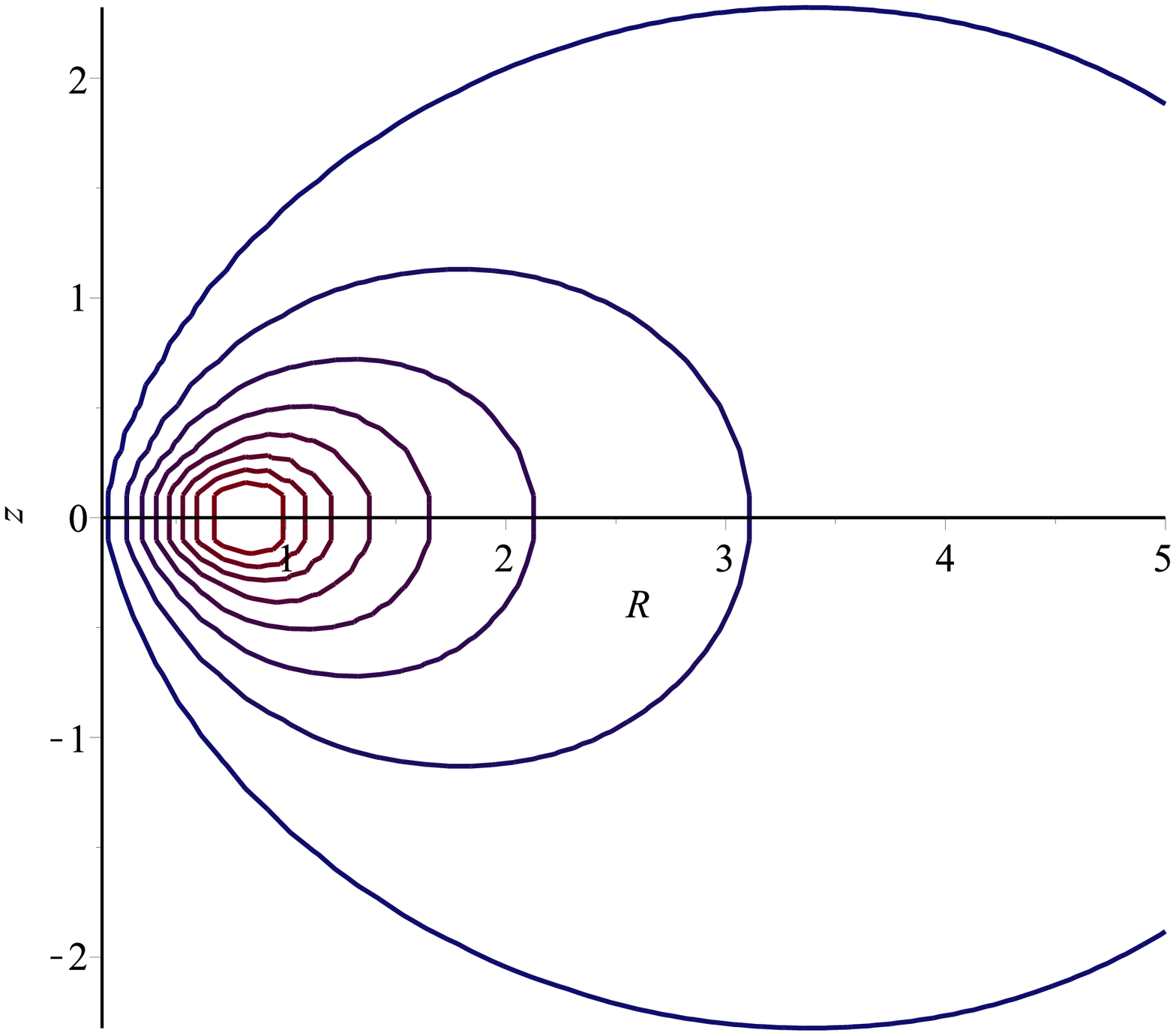}  
\end{array}
$$	
\caption{(a) Surface and  (b) level curves of the functions $RA$ as functions of $R$ and
$z$ for $b_1 = 0.5$.}
\label{fig:lineas}
\end{figure}

In Fig. \ref{fig:lineas} we have  plotted  the surface and level curves of the function
$RA$ that  represents
the  magnetic field lines as a function of $R$ and $z$ for the  parameter  $b_1 =
0.5$. We find that
the lines  are closed curves  surrounding  the disk which suggests
that the source  of the magnetic  field is a  ringlike current distribution and
that the  field
is poloidal.

\section*{4. Conclusions}

In the present work we constructed   a family  of  finite thin disks  with  electric  current in presence of a poloidal 
magnetic field
by using  oblate spheroidal coordinates.  
These models permit to determine the magnetic field structure near to disk
knowing  the experimental  magnetic field strength   far away from the disk.  
The profile of the azimuthal  electric current density  on the disk surface  or the radial 
component of the magnetic field   also  need to be assumed.

The models are illustrated  for a particular profile  of   anular  azimuthal  electric current  density.
For these structures we analyse  the rotation curves  for
equatorial circular orbits of charged test particles both inside and outside the disk
and also its  stability against  radial perturbation
using a extension of the Rayleigh criterion.  We find that the presence of the magnetic field can affect in  different
ways the  movement of particles. 
 
%---------------------------------------------------------------------
%---------------------------------------------------------------------

%Bibliografia

\newpage

%---------------------------------------------------------------------
%---------------------------------------------------------------------

%figuras

%Corriente

%Rapidez

%Momentum angular

%lineas de campo magnetico

\end{document}